\documentclass[aps,twocolumn]{revtex4}

\usepackage{amsmath}
\usepackage{graphicx}

\usepackage{epsfig}

\def\ov#1{\overline{#1}}

\def\vb#1{\mbox{\boldmath$#1$}}
\def\pd#1#2{\frac{\partial #1}{\partial #2}}

\def\wh#1{\widehat{#1}}
\def\bdot{\,\vb{\cdot}\,}
\def\btimes{\,\vb{\times}\,}

\def\bhat{\wh{{\sf b}}}
\def\exd{{\sf d}}
\def\cal#1{\mathcal{#1}}

\newcommand{\bc}{\begin{center}}
\newcommand{\ec}{\end{center}}
\newcommand{\bt}{\begin{tabbing}}
\newcommand{\et}{\end{tabbing}} 
\newcommand{\be}{\begin{eqnarray*}}
\newcommand{\ee}{\end{eqnarray*}}
\newcommand{\bs}{\begin{slide}}
\newcommand{\es}{\end{slide}}

\begin{document}

\title{Canonical transformation for trapped/passing guiding-center orbits \\ in axisymmetric tokamak geometry}

\author{Alain J.~Brizard$^{1}$ and Fran\c{c}ois-Xavier Duthoit$^{2,3}$}
\affiliation{$^{1}$Department of Physics, Saint Michael's College Colchester, VT 05439, USA \\ $^{2}$IRFM, CEA, F-13108, Saint-Paul-lez-Durance, France \\
$^{3}$ Department of Nuclear Engineering, Seoul National University, Seoul 151-742, South Korea} 

\begin{abstract}
The generating function for the canonical transformation from the parallel canonical coordinates $(s,p_{\|})$ to the action-angle coordinates 
$(\zeta,J)$ for trapped/passing guiding-center orbits in axisymmetric tokamak geometry is presented. Drawing on the analogy between the phase-space portraits of the librating/rotating pendulum and the trapped/passing guiding-center orbits, the generating function is expressed in terms of the Jacobi zeta function, which can then readily be used to obtain an explicit expression for the bounce-center transformation for trapped/passing-particle guiding-center orbits in axisymmetric tokamak geometry. 
\end{abstract}


\maketitle

\section{Introduction}

Magnetically-confined trapped-particle and passing-particle guiding-center orbits are the hallmark of axisymmetric tokamak geometry 
\cite{RBW,Cary_Brizard}. In a recent paper \cite{Brizard_gc_tok}, the mathematical representation of the poloidal angle $\vartheta$ and the parallel guiding-center momentum $p_{\|}$ for these orbits in simple axisymmetric tokamak geometry  (i.e., with circular concentric magnetic surfaces \cite{RBW}) was given in terms of the Jacobi elliptic functions and complete elliptic integrals \cite{NIST_Jacobi,NIST_elliptic,Lawden}, expressed in terms of the action-angle coordinates $(\zeta,J)$ for the bounce and transit orbits.

The purpose of the present paper is to derive an explicit expression for the generating function $S(\zeta,J)$ for the canonical transformation from the parallel coordinates $(s,p_{\|})$ to the action-angle coordinates $(\zeta, J)$. These two canonical pairs satisfy the canonical relation \cite{Goldstein}
\begin{equation}
\pd{s}{\zeta}\,\pd{p_{\|}}{J} - \pd{s}{J}\,\pd{p_{\|}}{\zeta} \;=\; 1,
\label{eq:ps_Jzeta_RGL}
\end{equation}
which plays a crucial role in the bounce-center phase-space transformation originally described by Littlejohn \cite{RGL_82} in general magnetic geometry. Littlejohn describes Eq.~\eqref{eq:ps_Jzeta_RGL} as an awkward relation to prove \cite{RGL_footnote}, however, because the parallel coordinates 
$s(\zeta,J)$ and $p_{\|}(\zeta,J)$ are often unknown functions of the action-angle coordinates in general magnetic geometry. The bounce-center transformation was therefore only formally known until recently \cite{Brizard_gc_tok} when the parallel coordinates were explicitly expressed in terms of Jacobi elliptic functions for simple axisymmetric tokamak geometry.

The generating function $S(\zeta,J)$ for the canonical transformation $(s,p_{\|}) \rightarrow (\zeta,J)$ is defined by the one-form identity 
\cite{Goldstein}
\begin{equation}
p_{\|}\,\exd s \;\equiv\; J\,\exd\zeta \;+\; \exd S,
\label{eq:dS_eq}
\end{equation}
which yields the two-form relation $\exd p_{\|}\,\wedge\,\exd s = \exd J\,\wedge\,\exd\zeta$ from which Eq.~\eqref{eq:ps_Jzeta_RGL} follows. In simple axisymmetric tokamak geometry, the magnetic field is ${\bf B} \equiv \nabla\xi\btimes\nabla\psi$, where the magnetic-field-line labels 
$(\psi,\xi)$ are the poloidal magnetic flux $\psi$ and the Euler potential $\xi \equiv \varphi - q(\psi)\,\vartheta$, which is expressed in terms of the toroidal and poloidal angles $\varphi$ and $\vartheta$, with the safety factor $q(\psi) \equiv {\bf B}\bdot\nabla\varphi/({\bf B}\bdot\nabla\vartheta)$ assumed to be a function of the poloidal magnetic flux only. Since the unit magnetic-field vector $\bhat \equiv b_{\varphi}\,\nabla\varphi + b_{\vartheta}\,\nabla\vartheta$ has only toroidal and poloidal components in simple axisymmetric tokamak geometry, we easily obtain the expression 
\begin{eqnarray}
\bhat\bdot\exd{\bf X} & = & b_{\varphi}\,\exd\varphi + b_{\vartheta}\,\exd\vartheta \;\equiv\; R_{\|}\;\exd\vartheta + b_{\varphi}\,(\exd\xi + \vartheta\,\exd q) \nonumber \\
 & \equiv & \exd s \;+\; {\cal R}_{\xi}\,\exd\xi \;+\; {\cal R}_{\psi}\,\exd\psi,
\label{eq:R||_def}
\end{eqnarray}
where we used $\exd\varphi = q\,\exd\vartheta + (\exd\xi + \vartheta\,q^{\prime}\exd\psi)$, we defined the connection length $R_{\|} \equiv q\,
b_{\varphi} + b_{\vartheta}$, and the magnetic-field labels $y^{a} = (\psi,\xi)$ satisfy the relations \cite{Brizard_2000} $\bhat\bdot\nabla y^{a} \equiv 0$ and ${\cal R}_{a} \equiv \bhat\bdot\partial{\bf X}/\partial y^{a} = (\vartheta\,b_{\varphi}\,q^{\prime},\;b_{\varphi})$. Hence, by holding the magnetic-field labels $(\psi,\xi)$ constant, we arrive at the lowest-order definition $\exd s \equiv R_{\|}\,\exd\vartheta$. From the identity 
\eqref{eq:dS_eq}, we therefore obtain the differential equations
\begin{eqnarray}
\pd{S}{\zeta} & = & p_{\|}\,R_{\|}\;\pd{\vartheta}{\zeta} \;-\; J, \label{eq:S_zeta} \\
\pd{S}{J} & = & p_{\|}\,R_{\|}\;\pd{\vartheta}{J}, \label{eq:S_J}
\end{eqnarray}
whose solutions require the knowledge of the parallel coordinates $\vartheta(\zeta,J)$ and $p_{\|}(\zeta,J)$ derived in 
Ref.~\cite{Brizard_gc_tok}. By integrating Eq.~\eqref{eq:S_zeta} with respect to $\zeta$, we obtain the generating function 
\begin{equation}
S(\zeta,J) \;=\; \int_{0}^{\zeta}p_{\|}\,R_{\|}\;\pd{\vartheta}{\zeta^{\prime}}\;d\zeta^{\prime} \;-\; J\,\zeta, 
\label{eq:S_sol}
\end{equation}
which must then be subject to the constraint \eqref{eq:S_J}.

The derivation of the generating function \eqref{eq:S_sol} allows the application of Lie-transform perturbation methods in the theory of bounce-center dynamics in axisymmetric tokamak geometry \cite{RGL_82}. The transformation from the magnetic-field labels $y^{a} = (\psi,\xi)$ to the bounce-center magnetic-field labels $Y^{a} = (\Psi, \Xi)$ is obtained by Lie-transform perturbation methods \cite{Cary_Brizard,Brizard_2000} up to first order as
\begin{equation}
Y^{a} - y^{a} \;=\; \varepsilon_{\rm gc}\;G_{1}^{a} \equiv -\;\varepsilon_{\rm gc}\,\frac{c}{e}\,\eta^{ab}\left(\pd{S}{y^{b}} + p_{\|}\,{\cal R}_{b} \right),
\label{eq:G1_bc}
\end{equation}
where $\eta^{12} = 1 = -\,\eta^{21}$ and $\varepsilon_{\rm gc}$ denotes the guiding-center ordering parameter 
\cite{Cary_Brizard}. Hence, once the generating function \eqref{eq:S_sol} is obtained, the bounce-center transformation \eqref{eq:G1_bc} can be constructed at first order (see Sec.~\ref{sec:bc} for additional details).

The remainder of the paper is organized as follows. In Sec.~\ref{sec:gc_tok}, we review the mathematical representation of the poloidal angle $\vartheta$ and the parallel guiding-center momentum $p_{\|}$ for guiding-center orbits in simple axisymmetric tokamak geometry given in terms of the Jacobi elliptic functions $({\rm cn}, {\rm sn}, {\rm dn})$ in Ref.~\cite{Brizard_gc_tok}. In Sec.~\ref{sec:S_def}, the generating function $S(\zeta,J)$ is derived for each class of guiding-center orbits and is expressed in terms of the Jacobi zeta function \cite{NIST_Jacobi,Lawden}, in complete analogy with the planar-pendulum problem discussed in Ref.~\cite{Brizard_CNSNS}. In Sec.~\ref{sec:bc}, we present the bounce-center transformation \eqref{eq:G1_bc} in axisymmetric tokamak geometry up to first order. Lastly, we summarize our work in Sec.~\ref{sec:sum} and present mathematical details in Appendices A and B associated with the proofs that our solutions for the generating function $S(\zeta,J)$ satisfy the relations \eqref{eq:S_zeta}-\eqref{eq:S_J}.

\section{\label{sec:gc_tok}Guiding-center Orbits in Simple Tokamak Geometry}

The guiding-center equations of motion for the parallel coordinates $(s,p_{\|})$ were recently solved \cite{Brizard_gc_tok} in terms of the Jacobi elliptic functions $({\rm cn},{\rm sn}, {\rm dn})$ in simple axisymmetric tokamak geometry. Here, the magnitude of the magnetic field is
\begin{eqnarray}
B(\vartheta; \psi) & \simeq & B_{0} \left( 1 \;-\frac{}{} \epsilon\,\cos\vartheta \right) \nonumber \\
 & = & B_{\rm e} \;+\; 2B_{0}\epsilon\;
\sin^{2}\frac{\vartheta}{2},
\label{eq:B_simple}
\end{eqnarray}
where $\epsilon(\psi) \equiv r(\psi)/R \ll 1$ denotes the small inverse aspect ratio ($R$ denotes the major radius of the magnetic axis, located at 
$\epsilon = 0$) and $B_{\rm e}(\psi) \equiv B_{0}\,(1 - \epsilon)$ denotes the magnitude of the magnetic field on the outside equatorial plane 
(at $\vartheta= 0$). 

The total energy of a guiding-center particle (of mass $m$ and charge $e$)
\begin{equation}
{\cal E} \;=\; \mu\,B_{\rm e} \;+\; \left( \frac{p_{\|}^{2}}{2m} \;+\; 2\epsilon\;\mu\,B_{0}\;\sin^{2}\frac{\vartheta}{2} \right) 
\label{eq:gc_energy}
\end{equation}
is a function of the parallel variables $(\vartheta, p_{\|})$, the guiding-center magnetic moment $\mu$, and the poloidal magnetic flux $\psi$. For each invariant pair $({\cal E}, \mu)$, the guiding-center parallel momentum is a function of the poloidal angle
\begin{equation}
p_{\|}(\vartheta) \;=\; 2\,\sigma\;\sqrt{\epsilon\,m\;\mu\,B_{0} \left( \kappa \;-\; \sin^{2}\frac{\vartheta}{2} \right)},
\label{eq:p||_theta}
\end{equation}
where $\sigma = \pm\,1$ denotes the sign of $p_{\|}$ and the dimensionless bounce-transit parameter is \cite{footnote_1}
\begin{equation}
\kappa({\cal E}, \mu, \psi) \;\equiv\; \frac{{\cal E} - \mu\,B_{\rm e}}{2\epsilon\;\mu B_{0}}.
\label{eq:kappa}
\end{equation}
The trapped/passing-particle guiding-center orbits are parameterized by $\kappa$: trapped-particle orbits are defined by $0 < \kappa < 1$ (the turning points of $p_{\|}$ are at $\vartheta_{\rm b} = \pm\,2\,\arcsin\sqrt{\kappa}$), passing-particle orbits are defined by $\kappa > 1$, and the trapped-passing boundary is $\kappa = 1$. In what follows, we will use the definitions
\begin{eqnarray}
p_{\|{\rm e}} & = & \sqrt{2m\,({\cal E} - \mu\,B_{\rm e})} \;=\; 2\sqrt{\kappa}\;m\,\omega_{\|}\,R_{\|}, \label{eq:p||e_def} \\
J_{\|} & = & m\,\omega_{\|}\,R_{\|}^{2} \;=\; p_{\|{\rm e}} R_{\|}/(2\sqrt{\kappa}), \label{eq:J||_def}
\end{eqnarray}
where $\omega_{\|}R_{\|} \equiv \sqrt{\epsilon\;\mu B_{0}/m}$. Here, $p_{\|{\rm e}}$ denotes the maximum parallel momentum on the equatorial plane 
(at $\vartheta = 0$) while $J_{\|}$ characterizes the parametric dependence of the bounce and transit actions. 

Lastly, we note that, in axisymmetric tokamak geometry, the guiding-center toroidal canonical angular momentum
\begin{equation}
P_{\varphi} \;\equiv\; -\,\frac{e}{\varepsilon_{\rm gc}\,c}\;\psi \;+\; p_{\|}\,b_{\varphi} \;\equiv\; -\,\frac{e}{\varepsilon_{\rm gc}\,c}\;\ov{\psi}
\label{eq:Pphi_def}
\end{equation}
is an exact invariant for the trapped/passing-particle guiding-center orbits. Hence, to lowest order in $\varepsilon_{\rm gc}$, the magnetic flux $\psi \equiv \ov{\psi}$ is constant in Eqs.~\eqref{eq:kappa}-\eqref{eq:J||_def}. Magnetic drifts from this constant magnetic surface are generated by $p_{\|}\,
b_{\varphi}$ and are discussed in Sec.~\ref{subsec:mag}.

\subsection{Jacobi representation of trapped-particle orbits}

The bounce action associated with a trapped-particle orbit is \cite{Wang_Hahm,Brizard_gc_tok}
\begin{equation}
J_{\rm b} \;=\; \frac{8J_{\|}}{\pi} \;\left[ {\sf E}(\kappa) \;-\frac{}{} (1 - \kappa)\;{\sf K}(\kappa) \right],
\label{eq:J_bounce}
\end{equation}
where ${\sf K}$ and ${\sf E}$ denote the complete elliptic integrals of the first and second kind \cite{NIST_elliptic}. The bounce frequency is defined from Eq.~\eqref{eq:J_bounce} as
\begin{equation}
\omega_{\rm b} \;\equiv\; (\partial J_{\rm b}/\partial{\cal E})^{-1} \;=\; \nu_{\rm b}(\kappa)\;\omega_{\|},
\label{eq:omega_b}
\end{equation}
where the bounce coefficient
\begin{equation}
\nu_{\rm b}(\kappa) \;\equiv\; \frac{\pi}{2\,{\sf K}(\kappa)},
\label{eq:nu_b_def}
\end{equation}
varies from $\nu_{\rm b}(0) = 1$ (deeply-trapped limit) to $\nu_{\rm b}(1) = 0$ (separatrix limit). Hence, $\omega_{\|}$ denotes the bounce frequency of deeply-trapped particles.

The solutions for the trapped-particle ($\kappa < 1$) guiding-center orbits are expressed in terms of the pendulum-like solutions \cite{footnote_2}
\begin{eqnarray}
p_{\|} & = & p_{\|{\rm e}}\;{\rm cn}(\chi_{\rm b}|\kappa), \label{eq:p_b} \\
\vartheta & = & 2\;\arcsin\left[ \sqrt{\kappa}\frac{}{}{\rm sn}(\chi_{\rm b}|\kappa)\right], \label{eq:theta_b}
\end{eqnarray}
where the bounce angle $\zeta_{\rm b}$ appears through the variable $\chi_{\rm b} \equiv \zeta_{\rm b}/\nu_{\rm b}$. 

\subsection{Jacobi representation of passing-particle orbits}

The transit action associated with a passing-particle orbit is \cite{Wang_Hahm,Brizard_gc_tok}
\begin{equation}
J_{\rm t} \;=\; \frac{4J_{\|}}{\pi}\;\sqrt{\kappa}\,{\sf E}(\kappa^{-1}).
\label{eq:J_transit}
\end{equation}
The transit frequency is thus expressed as
\begin{equation}
\omega_{\rm t} \;\equiv\; (\partial J_{\rm t}/\partial{\cal E})^{-1} \;=\; \nu_{\rm t}(\kappa)\;\omega_{\|},
\label{eq:omega_t}
\end{equation}
where the transit coefficient
\begin{equation}
\nu_{\rm t}(\kappa) \;\equiv\; \frac{\pi\,\sqrt{\kappa}}{{\sf K}(\kappa^{-1})},
\label{eq:nu_t_def}
\end{equation}
varies from $\nu_{\rm t}(1) = 0$ (separatrix limit) to $\nu_{\rm t}(\kappa) \simeq 2\,\sqrt{\kappa}$ as $\kappa \gg 1$ (strongly-circulating limit). 

According to Ref.~\cite{Brizard_gc_tok}, the solutions for the passing-particle ($\kappa > 1$) guiding-center orbits are expressed in terms of the pendulum-like solutions \cite{Brizard_CNSNS}
\begin{eqnarray}
p_{\|} & = & p_{\|{\rm e}}\;{\rm dn}(\chi_{\rm t}|\kappa^{-1}), \label{eq:p_dn} \\
\vartheta & = & 2\;\arcsin\left[ \frac{}{}{\rm sn}(\chi_{\rm t}|\kappa^{-1})\right], \label{eq:theta_sn}
\end{eqnarray}
where the transit angle $\zeta_{\rm t}$ appears through the variable $\chi_{\rm t} \equiv \sqrt{\kappa}\,\zeta_{\rm t}/\nu_{\rm t}$. We note that the period of the Jacobi elliptic function ${\rm dn}(\chi_{\rm t}|\kappa^{-1})$ is $2\,{\sf K}(\kappa^{-1})$.

The connection between Eqs.~\eqref{eq:J_bounce} and \eqref{eq:nu_b_def} for trapped particles and Eqs.~\eqref{eq:J_transit} and 
\eqref{eq:nu_t_def} for passing particles is expressed in terms of the identities \cite{NIST_elliptic}
\begin{eqnarray}
{\rm Re}[{\sf K}(\kappa)] & = & {\sf K}(\kappa^{-1})/\sqrt{\kappa}, \\
{\rm Re}[{\sf E}(\kappa)] & = & \sqrt{\kappa}\;\left[{\sf E}(\kappa^{-1}) - (1 -\kappa^{-1})\,{\sf K}(\kappa^{-1})\right],
\end{eqnarray}
and the missing factor of 2 in Eqs.~\eqref{eq:J_transit} and \eqref{eq:nu_t_def} comes from the definition of the transit action. The connection between the trapped-particle solutions \eqref{eq:p_b}-\eqref{eq:theta_b} and the passing-particle solutions \eqref{eq:p_dn}-\eqref{eq:theta_sn} is expressed in terms of the identities (where $\kappa > 1$ on the left side)
\begin{equation}
\left. \begin{array}{rcl}
\sqrt{\kappa}\;{\rm sn}(t|\kappa) & = & {\rm sn}\left(\sqrt{\kappa}\,t|\kappa^{-1}\right) \\
 &  & \\
{\rm cn}(t|\kappa) & = & {\rm dn}\left(\sqrt{\kappa}\,t|\kappa^{-1}\right) \\
 &  & \\
{\rm dn}(t|\kappa) & = & {\rm cn}\left(\sqrt{\kappa}\,t|\kappa^{-1}\right)
\end{array} \right\}.
\label{eq:connection}
\end{equation}
The Jacobi elliptic functions ${\rm sn}$ and ${\rm cn}$ on the right of Eq.~\eqref{eq:connection} are periodic with quarter period ${\sf K}(\kappa^{-1})/\sqrt{\kappa}$, while ${\rm dn}$ is periodic with half period ${\sf K}(\kappa^{-1})/\sqrt{\kappa}$.

\subsection{\label{subsec:mag}Magnetic-drift surfaces}

The formulas \eqref{eq:p_b}-\eqref{eq:theta_b} and \eqref{eq:p_dn}-\eqref{eq:theta_sn} for the parallel coordinates $(p_{\|},\vartheta)$ are derived under the assumption that the poloidal magnetic flux $\psi$ is held constant in Eqs.~\eqref{eq:kappa}-\eqref{eq:J||_def}. 

\begin{figure}
\epsfysize=2.5in
\epsfbox{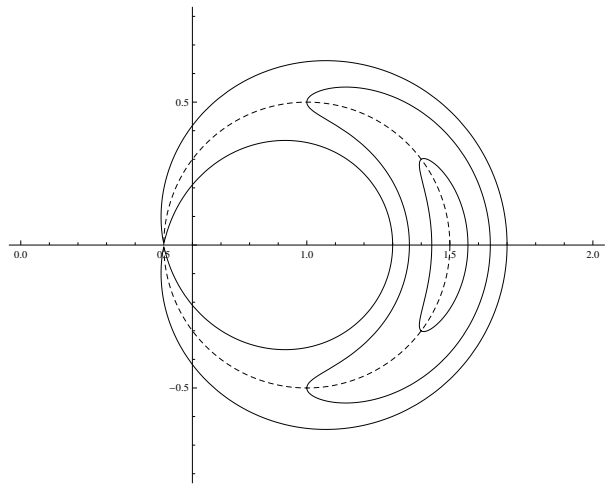}
\caption{Trapped-particle orbits on the poloidal-plane for $\kappa = 0.1$ (deeply-trapped), 0.5, and 0.9999 (barely-trapped). The magnetic surface labeled by the invariant flux $\ov{\psi}$ is shown as a dashed circle. (See Ref.~\cite{Brizard_gc_tok} for details.)}
\label{fig:banana}
\end{figure}

\begin{figure}
\epsfysize=2.5in
\epsfbox{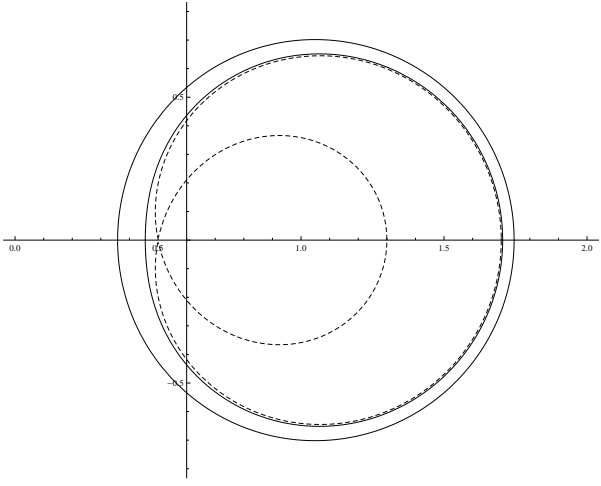}
\caption{Passing-particle orbits on the poloidal-plane for $\kappa = 1.05$ (barely-passing) and 1.5 (passing). The trapped-passing boundary ($\kappa = 
1$) is shown as a dashed curve. (See Ref.~\cite{Brizard_gc_tok} for details.)}
\label{fig:passing}
\end{figure}

For trapped-particle guiding-center orbits in axisymmetric tokamak geometry, the poloidal magnetic flux at the bounce points (where $p_{\|} = 0$) 
$\ov{\psi} \equiv -\,(c/e)\,P_{\varphi}$ is simply related to the toroidal canonical angular momentum \eqref{eq:Pphi_def}, which is an exact invariant for trapped-particle and passing-particle guiding-center orbits as a result of toroidal axisymmetry. Hence, as we follow a trapped-particle guiding-center orbit on the poloidal plane (at constant $\varphi$), the poloidal magnetic flux $\psi = \ov{\psi} + \varepsilon_{\rm gc}\,(c/e)\,p_{\|}\,\ov{b}_{\varphi} + \cdots$, which is obtained by inverting the relation \eqref{eq:Pphi_def}, traces a curve (solid lines in Fig.~\ref{fig:banana}) that lies on a magnetic-drift surface labeled by the guiding-center invariants $(\ov{\psi},{\cal E},\mu)$. Figure \ref{fig:passing} shows that a less-pronounced magnetic-drift motion takes place in the poloidal plane for passing-particle guiding-center orbits.

The magnetic-field label $\xi = \varphi - q(\psi)\,\vartheta$ also changes in the course of the bounce motion of a trapped-particle. The bounce-averaged drift-precession frequency 
$\omega_{\rm d} \equiv \langle\dot{\xi}\rangle_{\rm b}$ can be calculated as \cite{RBW,Brizard_gc_tok}
\begin{equation}
\omega_{\rm d} \;=\; -\;\frac{c}{e}\,\mu\;\left\langle \pd{B}{\psi}\right\rangle_{\rm b} \;=\; \Omega_{0}\;\frac{q\,\rho_{0}^{2}}{\epsilon\;R^{2}}
\left( \frac{\sf E}{\sf K} - \frac{1}{2}\right),
\label{eq:omega_d}
\end{equation}
where $\Omega_{0}$ and $\rho_{0} \equiv \sqrt{2\,\mu\,B_{0}/m\Omega_{0}^{2}}$ denote the gyrofrequency and gyroradius evaluated at the magnetic axis. Corrections to Eq.~\eqref{eq:omega_d} due to magnetic shear $(q^{\prime} \neq 0)$ and shifted (non-concentric) magnetic surfaces are considered, for example, in Ref.~\cite{Wu_CZC_RBW}.

\section{\label{sec:S_def}Generating Functions}

We are now able to find expressions for the generating functions $S(\zeta,J)$ that generate the canonical transformation $(s,p_{\|}) \rightarrow 
(\zeta,J)$ for trapped-particle and passing-particle guiding-center orbits in axisymmetric tokamak geometry. 

Because the phase portrait $(\vartheta,\dot{\vartheta})$ for the trapped-particle/passing-particle orbits is analogous to the libration/rotation motion of a planar pendulum \cite{Brizard_gc_tok}, we borrow from the analysis of the pendulum problem \cite{Brizard_CNSNS}, which reveals that the generating function for the trapped-particle/passing-particle canonical transformation is again be expressed in terms of the Jacobi zeta function.

\subsection{Trapped-particle orbits}

\begin{figure}
\epsfysize=2in
\epsfbox{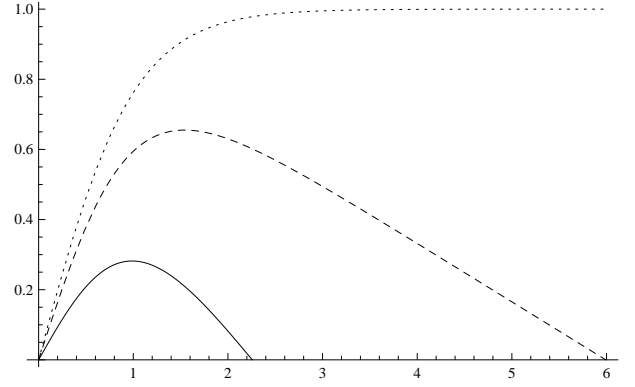}
\caption{Plots of ${\rm Z}(\chi|\kappa)$ in the range $0 < \chi < {\sf K}$ for $\kappa = 0.8$ (solid) and $\kappa = 0.9999$ (dashed). The separatrix limit ${\rm Z}(\chi|\kappa = 1) = \tanh(\chi)$ is also shown (dot-dashed).}
\label{fig:zeta}
\end{figure}

We first derive the generating function $S_{\rm b}(\zeta_{\rm b},J_{\rm b})$ for the canonical transformation $(s,p_{\|}) \rightarrow (\zeta_{\rm b},
J_{\rm b})$ for the trapped-particle guiding-center orbits. From the trapped-particle solutions \eqref{eq:p_b}-\eqref{eq:theta_b}, we obtain
\begin{equation}
p_{\|}\,R_{\|}\;\pd{\vartheta}{\chi_{\rm b}} \;=\; 4J_{\|}\;\kappa\,{\rm cn}^{2}(\chi_{\rm b}|\kappa),
\label{eq:ps_zeta}
\end{equation}
where $\chi_{\rm b} = \zeta_{\rm b}/\nu_{\rm b}$, so that Eq.~\eqref{eq:S_sol} becomes
\begin{eqnarray}
S_{\rm b} & = & 4\,J_{\|} \left[ \kappa\; \int_{0}^{\chi_{\rm b}}{\rm cn}^{2}(u|\kappa)\,du \right] \;-\; J_{\rm b}\,\zeta_{\rm b}
\nonumber \\
 & \equiv & 4\,J_{\|}\;{\rm Z}(\chi_{\rm b}|\kappa).
\label{eq:S_Z_b}
\end{eqnarray}
In Eq.~\eqref{eq:S_Z_b}, the Jacobi zeta function \cite{NIST_Jacobi,Lawden}
\begin{equation}
{\rm Z}(\chi|\kappa) \;\equiv \int_{0}^{\chi}\left({\rm dn}^{2}(u|\kappa) \;-\; \frac{\sf E}{\sf K}\right)\,du
\label{eq:Z_def}
\end{equation}
is an odd function of $\chi$: ${\rm Z}(-\,\chi|\kappa) = -\,{\rm Z}(\chi|\kappa)$, it has a period of $2{\sf K}$: ${\rm Z}(\chi + 2{\sf K}|\kappa) = 
{\rm Z}(\chi|\kappa)$, and it vanishes at $n\,{\sf K}$: ${\rm Z}(n\,{\sf K}|\kappa) = 0$ for $n = 0, \pm 1, ...$. The Jacobi zeta function 
\eqref{eq:Z_def} is shown in Fig.~\ref{fig:zeta} for $\kappa = 0.8$ (solid) and $\kappa = 0.9999$ (dashed), in addition to the separatrix limit 
${\rm Z}(\chi|\kappa = 1) = \int_{0}^{\chi}\,{\rm sech}^{2}u\,du = \tanh\chi$. 

Note that the zeta function ${\rm Z}(\chi|\kappa)$ can also be defined in terms of the Jacobi theta function $\theta_{4}(\zeta|\tau)$ as \cite{Lawden}
\begin{equation}
{\rm Z}(\chi_{\rm b}|\kappa) \;\equiv\; \nu_{\rm b}(\kappa)\;\pd{\ln\theta_{4}(\zeta_{\rm b}|\tau)}{\zeta_{\rm b}},
\label{eq:Z_theta_def}
\end{equation}
where the parameter $\tau(\kappa) \equiv i\,{\sf K}(1 - \kappa)/{\sf K}(\kappa)$ and the theta functions $\theta_{n}$ $(n=1,2,3,4)$ satisfy the ``heat'' equation \cite{Lawden}
\begin{equation}
i\;\pd{\theta_{n}}{\tau} \;=\; -\,\frac{\pi}{4}\;\frac{\partial^{2}\theta_{n}}{\partial\zeta^{2}}.
\label{eq:theta_heat}
\end{equation}
Hence, Eq.~\eqref{eq:S_Z_b} can be expressed as
\begin{equation}
S_{\rm b} \;=\; 4\,J_{\|}\,\nu_{\rm b}\;\pd{\ln\theta_{4}(\zeta_{\rm b}|\tau)}{\zeta_{\rm b}},
\label{eq:S_theta4}
\end{equation}
which clearly shows that the generating function $S_{\rm b}$ is explicitly bounce-angle-dependent. In App.~\ref{sec:proof}, Eq.~\eqref{eq:S_Z_b} [or \eqref{eq:S_theta4}] is shown to satisfy the relations \eqref{eq:S_zeta}-\eqref{eq:S_J}.

\subsection{Passing-particle orbits}

Next, we derive the generating function $S_{\rm t}(\zeta_{\rm t},J_{\rm t})$ for the canonical transformation $(s,p_{\|}) \rightarrow (\zeta_{\rm t},
J_{\rm t})$ for the passing-particle guiding-center orbits. From the passing-particle solutions \eqref{eq:p_dn}-\eqref{eq:theta_sn}, we obtain
\begin{equation}
p_{\|}\,R_{\|}\;\pd{\vartheta}{\chi_{\rm t}} \;=\; 4J_{\|}\,\sqrt{\kappa}\;\ov{\rm dn}^{2},
\label{eq:ps_zeta_t}
\end{equation}
where $\chi_{\rm t} = \sqrt{\kappa}\,\zeta_{\rm t}/\nu_{\rm t}$ and we use the notation $\ov{\rm pq} = {\rm pq}(\chi_{\rm t}|\kappa^{-1})$, so that 
Eq.~\eqref{eq:S_sol} becomes
\begin{eqnarray}
S_{\rm t} & = & 4\,J_{\|}\;\sqrt{\kappa}\,\int_{0}^{\chi_{\rm t}}\;{\rm dn}^{2}(u|\kappa^{-1})\,du \;-\; J_{\rm t}\,\zeta_{\rm t} \nonumber \\
 & \equiv & 4\,J_{\|}\;\sqrt{\kappa}\;{\rm Z}(\chi_{\rm t}|\kappa^{-1}).
\label{eq:S_Z_t}
\end{eqnarray}
Here, we used the definition \eqref{eq:Z_def}:
\begin{equation}
\int_{0}^{\chi_{\rm t}}\,\ov{\rm dn}^{2}\,du \;=\; {\rm Z}(\chi_{\rm t}|\kappa^{-1}) \;+\; \frac{\ov{\sf E}}{\ov{\sf K}}\;\chi_{\rm t},
\end{equation}
and $4\,J_{\|}\,\sqrt{\kappa}\,(\ov{\sf E}/\ov{\sf K})\;\chi_{\rm t} \equiv J_{\rm t}\,\zeta_{\rm t}$. We may also write Eq.~\eqref{eq:S_Z_t} as
\begin{equation}
S_{\rm t} \;\equiv\; 4\,J_{\|}\nu_{\rm t}\;\pd{\ln\theta_{4}(\zeta_{\rm t}|\ov{\tau})}{\zeta_{\rm t}},
\label{eq:St_theta4}
\end{equation}
where $\ov{\tau} = i{\sf K}(1 - \kappa^{-1})/{\sf K}(\kappa^{-1})$. In App.~\ref{sec:proof}, we show that the solution \eqref{eq:S_Z_t} [or 
\eqref{eq:St_theta4}] satisfies the relations \eqref{eq:S_zeta}-\eqref{eq:S_J}.

\section{\label{sec:bc}Bounce-center Transformation}

The generating function for the canonical transformation from the particle parallel coordinates $(s,p_{\|})$ to the trapped/passing bounce-center action-angle coordinates $(\zeta,J)$ is constructed in Sec.~\ref{sec:S_def} as
\begin{equation}
S(\zeta, J) \;=\; 4\,J_{\|}\; \left\{ \begin{array}{lr}
{\rm Z}(\zeta_{\rm b}/\nu_{\rm b}|\kappa) \;\;\;\; & (\kappa < 1) \\
 & \\
\sqrt{\kappa}\;{\rm Z}(\sqrt{\kappa}\,\zeta_{\rm t}/\nu_{\rm t}|\kappa^{-1}) \;\;\;\; & (\kappa > 1)
\end{array} \right.
\label{eq:S_summary}
\end{equation}
where the trapped-particle case $(\kappa < 1)$ is defined in terms of the bounce coefficient $\nu_{\rm b}(\kappa) = \pi/2{\sf K}(\kappa)$ and the passing-particle case $(\kappa > 1)$ is defined in terms of the transit coefficient $\nu_{\rm t}(\kappa) = \pi\,\sqrt{\kappa}/{\sf K}(\kappa^{-1})$.

We can now proceed with the bounce-center transformation $y^{a} = (\psi,\xi) \rightarrow Y^{a} = (\Psi, \Xi)$ defined by \eqref{eq:G1_bc}, which applies to both trapped-particle and passing-particle guiding-center orbits.

\subsection{Bounce-center transformation $\psi \rightarrow \Psi$}

We first consider the bounce-center transformation $\psi \rightarrow \Psi$ generated to first order in $\varepsilon_{\rm gc}$ by
\begin{equation}
G_{1}^{\psi} \;=\; -\;\frac{c}{e} \left( \pd{S}{\xi} \;+\; p_{\|}\,{\cal R}_{\xi} \right) \;=\; -\;\frac{c}{e}\,p_{\|}\,b_{\varphi},
\label{eq:G1_psi}
\end{equation}
where $\partial S/\partial\xi = \partial S/\partial\varphi \equiv 0$ follows from the assumption of axisymmetric tokamak geometry and ${\cal R}_{\xi} = 
b_{\varphi}$ was used. Hence, the bounce-center magnetic-flux label
\begin{equation}
\Psi \;=\; \psi \;+\; \varepsilon_{\rm gc}\;G_{1}^{\psi} \;\equiv\; \ov{\psi} 
\label{eq:Psi_def}
\end{equation}
is defined as the canonical magnetic-flux invariant \eqref{eq:Pphi_def}. The physical interpretation of $G_{1}^{\psi}$ for a trapped/passing-particle guiding-center orbit is, therefore, that of a $\zeta$-dependent bounce-radius analogous to the gyroangle-dependent gyroradius in guiding-center theory \cite{Cary_Brizard}. 

This analogy is exact for trapped particles since the bounce average of the guiding-center parallel momentum \eqref{eq:p_b} vanishes:
\[ \langle p_{\|}\rangle_{\rm b} \;\equiv\; \frac{p_{\|{\rm e}}}{4\,{\sf K}}\int_{-2{\sf K}}^{2{\sf K}}{\rm cn}(u|\kappa)\,du \;=\; 0, \]
and, hence, the bounce-averaged magnetic flux $\langle\psi\rangle_{\rm b} = \ov{\psi}$ is the canonical magnetic-flux invariant $\ov{\psi}$ (to lowest order, i.e., treating $b_{\varphi} \simeq \ov{b}_{\varphi}$ as a constant), which is analogous to the guiding-center position corresponding to the gyro-averaged particle position (to lowest order, i.e., treating the magnetic field as uniform). 

The case of the transit average of the guiding-center parallel momentum \eqref{eq:p_dn} is different, however, since the transit average of the guiding-center parallel momentum \eqref{eq:p_dn} yields
\[ \langle p_{\|}\rangle_{\rm t} \;\equiv\; \frac{p_{\|{\rm e}}}{2\,\ov{\sf K}}\int_{0}^{2\ov{\sf K}}{\rm dn}(u|\kappa^{-1})\,du = m\,
\omega_{\rm t}R_{\|}, \]
where we used $\int_{0}^{2\ov{\sf K}}{\rm dn}(u|\kappa^{-1})\,du = \pi$ and the definitions \eqref{eq:p||e_def} and 
\eqref{eq:omega_t}-\eqref{eq:nu_t_def}. Hence, to lowest order (i.e., treating $b_{\varphi}$ as a constant), the transit-averaged magnetic flux $\langle
\psi\rangle_{\rm t} = \ov{\psi} + B_{\varphi}R_{\|}\,\omega_{\rm t}/\Omega$ is displaced from $\ov{\psi}$ by a shift 
\begin{equation}
B_{\varphi}R_{\|}\,\frac{\omega_{\rm t}}{\Omega} \;=\; \frac{q\nu_{\rm t}(\kappa)}{\sqrt{2\epsilon}}\;\rho_{0}|\nabla\psi|
\end{equation}
that vanishes at the trapped/passing boundary ($\kappa = 1$) and increases monotonically as a function of $\kappa > 1$, where $\rho_{0} = \sqrt{2\mu 
B_{0}/m\Omega^{2}}$ is the gyroradius on the magnetic axis.

Lastly, we note that, in guiding-center theory \cite{Brizard_2013}, the retention of higher-order corrections due to magnetic-field nonuniformity yields a connection between the gyro-averaged particle position and the guiding-center position that is related to guiding-center polarization effects. Likewise, the retention of higher-order corrections due to magnetic-field nonuniformity will lead to a bounce-averaged magnetic flux $\langle\psi
\rangle_{\rm b}$ that is no longer equal to the bounce-center magnetic-flux label $\Psi$, which will then lead to bounce-center polarization effects as discussed by Wang and Hahm \cite{Wang_Hahm}.

\subsection{Bounce-center transformation $\xi \rightarrow \Xi$}

Next, the bounce-center transformation $\xi \rightarrow \Xi$ is generated to first order in $\varepsilon_{\rm gc}$ by
\begin{equation}
G_{1}^{\xi} \;=\; \frac{c}{e} \left( \pd{S}{\psi} \;+\; p_{\|}\,b_{\varphi}\;\vartheta\,q^{\prime}\, \right),
\label{eq:G1_xi}
\end{equation}
where ${\cal R}_{\psi} = b_{\varphi}\;\vartheta\,q^{\prime}(\psi)$ was used. Hence, the bounce-center Euler potential $\Xi$ is defined up to first order as
\begin{eqnarray}
\Xi & = & \xi \;+\; \varepsilon_{\rm gc}\,G_{1}^{\xi} \nonumber \\
 & = & \left( \varphi \;+\; \varepsilon_{\rm gc}\,\frac{c}{e}\,\pd{S}{\psi} \right) \;-\; \vartheta \left[ q(\psi) \;+\; \varepsilon_{\rm gc}\,
G_{1}^{\psi}\;q^{\prime}(\psi)\right] \nonumber \\
& \equiv & \Phi \;-\; \vartheta(\zeta,J)\;q(\Psi),
\label{eq:Xi_def}
\end{eqnarray}
which includes a transformation of the toroidal angle $\Phi \equiv \varphi + \varepsilon_{\rm gc}\,G_{1}^{\varphi}$ generated by the canonical first-order component $G_{1}^{\varphi} = (c/e)\,\partial S/\partial\psi \equiv -\,\partial S/\partial P_{\varphi}$ (at lowest order in 
$\varepsilon_{\rm gc}$) and the safety factor $q(\Psi)$ in Eq.~\eqref{eq:Xi_def} is now evaluated at the bounce-center magnetic-flux label $\Psi \equiv \psi + \varepsilon_{\rm gc}\,G_{1}^{\psi}$ defined by Eq.~\eqref{eq:Psi_def}.

\section{\label{sec:sum}Summary}

The bounce-center transformation \eqref{eq:G1_bc} in axisymmetric tokamak geometry had until now only been solved in the deeply-trapped $(\kappa \ll 1)$ or the energetic-passing $(\kappa^{-1} \ll 1)$ limits (see Ref.~\cite{Wang_Hahm}, for example). By deriving the generating function \eqref{eq:S_summary} for the canonical transformation from the parallel guiding-center coordinates $(s,p_{\|})$ to the bounce-center action-angle coordinates $(\zeta,J)$ separately for the trapped-particle $(\kappa < 1)$ and the passing-particle $(\kappa > 1)$, we were able to find the bounce-center transformation 
\eqref{eq:Psi_def} and \eqref{eq:Xi_def} for all values of the bounce/transit parameter $\kappa$ up to first order. Higher-order corrections (including geometric corrections associated with realistic axisymmetric tokamak geometry) can also be calculated by Lie-transform methods.

\acknowledgments

Work by AJB was supported by a U.~S.~DoE grant under contract No.~DE-SC0006721. Work by FXD was supported by National R\& D Program through the National
Research Foundation of Korea (NRF) funded by the Ministry of Science, ICT \& Future Planning (No.~2013036120). This work, also supported by the
European Communities under the contract of association between EURATOM and CEA, was carried out within the framework of the European Fusion Development Agreement. The views and opinions expressed herein do not necessarily reflect those of the European Commission.

\appendix

\section{\label{sec:kappa_partial}$\kappa$-Derivatives of Jacobi Functions}

The $\kappa$-derivatives of the Jacobi elliptic functions $({\rm cn}, {\rm sn}, {\rm dn})$ are expressed in terms of the Jacobi zeta function 
\eqref{eq:Z_def} as
\begin{eqnarray}
\frac{\partial^{\prime}{\rm cn}}{\partial\kappa} & = & \frac{{\rm sn}\,{\rm dn}\;{\rm Z}}{2\,\kappa\,(1 - \kappa)} \;-\; 
\frac{{\rm sn}^{2}\,{\rm cn}}{2\,(1 - \kappa)}, 
\label{eq:cn_kappa} \\
\frac{\partial^{\prime}{\rm sn}}{\partial\kappa} & = & -\;\frac{{\rm cn}\,{\rm dn}\;{\rm Z}}{2\,\kappa\,(1 - \kappa)} \;+\; 
\frac{{\rm sn}\,{\rm cn}^{2}}{2\,(1 - \kappa)}, 
\label{eq:sn_kappa} \\
\frac{\partial^{\prime}{\rm dn}}{\partial\kappa} & = & \frac{{\rm cn}\,{\rm sn}\;{\rm Z}}{2\,(1 - \kappa)} \;-\; 
\frac{{\rm sn}^{2}\,{\rm dn}}{2\,(1 - \kappa)},
\label{eq:dn_kappa}
\end{eqnarray}
where the operator $\partial^{\prime}/\partial\kappa$ is defined as
\begin{equation}
\left.\pd{}{J}\right|_{\zeta} \;=\; \pd{\kappa}{J}\left( \pd{}{\kappa} \;+\; \left.\pd{\chi}{\kappa}\right|_{\zeta}\;\pd{}{\chi} \right) \;\equiv\;
\pd{\kappa}{J}\;\frac{\partial^{\prime}}{\partial\kappa},
\label{eq:J_zeta}
\end{equation}
with $\partial\kappa/\partial J \equiv \nu/(2\,J_{\|})$ for either trapped-particle $(\nu = \nu_{\rm b})$ or passing-particle $(\nu = \nu_{\rm t})$ solutions. The $\kappa$-derivative of the Jacobi zeta function, on the other hand, is expressed as
\begin{equation}
\frac{\partial^{\prime}{\rm Z}}{\partial\kappa} \;=\; -\;\frac{{\rm cn}^{2}\;{\rm Z}}{2\,(1 - \kappa)} \;+\; 
\frac{{\rm sn}\,{\rm cn}\,{\rm dn}}{2\,(1 - \kappa)}.
\label{eq:Z_kappa}
\end{equation}
Partial derivatives of elliptic functions $\ov{\rm pq} = {\rm pq}(\chi|\kappa^{-1})$ with respect to $\kappa$ are evaluated according to the chain rule
\[ \frac{\partial^{\prime}\ov{\rm pq}}{\partial\kappa} \;=\; -\;\frac{1}{\kappa^{2}} \left( \frac{\partial^{\prime}{\rm pq}}{\partial m}
\right)_{m = \kappa^{-1}}. \]

\section{\label{sec:proof}Generating Function}

In this Appendix, we show that the generating function \eqref{eq:S_summary} satisfies Eqs.~\eqref{eq:S_zeta}-\eqref{eq:S_J} for the trapped/passing-particle guiding-center orbits in simple axisymmetric tokamak geometry.

\subsection{Trapped-particle orbits}

Using Eq.~\eqref{eq:S_theta4}, we find
\begin{eqnarray}
\pd{S_{\rm b}}{\zeta_{\rm b}} & = & 4\,J_{\|}\,\nu_{\rm b} \left(\frac{\theta_{4}^{\prime}}{\theta_{4}}\right)^{\prime} \;=\; 4\,J_{\|}\,\nu_{\rm b} 
\left( \frac{\theta_{4}^{\prime\prime}}{\theta_{4}} \;-\; \frac{\theta_{4}^{\prime 2}}{\theta_{4}^{2}} \right) \nonumber \\
 & = & -\;\frac{8{\sf K}}{\pi}\,J_{\|}\,\left[ \left( \frac{{\sf E}}{{\sf K}} - 1\right) \;+\; \kappa\;{\rm sn}^{2} \right],
\label{eq:Sb_zeta_proof}
\end{eqnarray}
where we used (see Exercise 16 of Chapter 2 in Ref.~\cite{Lawden})
\[ \nu_{\rm b}\;\left(\frac{\theta_{4}^{\prime}}{\theta_{4}}\right)^{\prime} \;=\; \nu_{\rm b}\;\frac{\theta_{4}^{\prime\prime}(0)}{\theta_{4}(0)} \;-\;
\frac{\kappa}{\nu_{\rm b}}\;{\rm sn}^{2}, \]
and (formula 3.5.5 of Ref.~\cite{Lawden})
\[ \nu_{\rm b}\;\frac{\theta_{4}^{\prime\prime}(0)}{\theta_{4}(0)} \;=\; -\;\nu_{\rm b}^{-1}\left( \frac{\sf E}{\sf K} \;-\; 1 \right). \]
By using $p_{\|}R_{\|}\,\partial\vartheta/\partial\zeta_{\rm b}$, given by Eq.~\eqref{eq:ps_zeta}, we recover Eq.~\eqref{eq:S_zeta}
\begin{equation}
\pd{S_{\rm b}}{\zeta_{\rm b}} \;\equiv\; -\;J_{\rm b} \;+\; p_{\|}\,R_{\|}\;\pd{\vartheta}{\zeta_{\rm b}}.
\end{equation}
Using Eq.~\eqref{eq:S_Z_b}, we also recover Eq.~\eqref{eq:S_J}
\begin{eqnarray}
\pd{S_{\rm b}}{J_{\rm b}} & = & 2\,\nu_{\rm b}\;\frac{\partial^{\prime}{\rm Z}}{\partial\kappa} \;=\; \frac{\nu_{\rm b}}{(1 - \kappa)}\left(
{\rm sn}\,{\rm cn}\,{\rm dn} \;-\frac{}{} {\rm cn}^{2}\;{\rm Z}\right) \nonumber \\
 & = & \nu_{\rm b}\,{\rm cd} \left( {\rm sn} \;+\; 2\kappa\;\frac{\partial^{\prime}{\rm sn}}{\partial\kappa}\right) \;\equiv\; 
p_{\|}R_{\|}\;\pd{\vartheta}{J_{\rm b}},
\label{eq:S_J_proof}
\end{eqnarray}
where we used Eq.~\eqref{eq:sn_kappa}.

\subsection{Passing-particle orbits}

Using Eq.~\eqref{eq:St_theta4}, we find
\begin{eqnarray}
\pd{S_{\rm t}}{\zeta_{\rm t}} & = & 4\,\kappa\;J_{\|}\nu_{\rm t}\; \left( \frac{\ov{\theta}_{4}^{\prime\prime}}{\ov{\theta}_{4}} \;-\; 
\frac{\ov{\theta}_{4}^{\prime 2}}{\ov{\theta}_{4}^{2}} \right) \label{eq:St_zeta_proof} \\
 & = & -\;\frac{4}{\pi}\,J_{\|}\;\sqrt{\kappa}\,\ov{\sf E} \;+\; \frac{4}{\pi}\,J_{\|}\;\sqrt{\kappa}\,\ov{\sf K} \left( 1 \;-\; \kappa^{-1}\,
\ov{\rm sn}^{2} \right), \nonumber
\end{eqnarray}
where $\ov{\theta}_{4} \equiv \theta_{4}(\zeta_{\rm t}|\ov{\tau})$ and we used 
\[ \nu_{\rm t}\;\left(\frac{\ov{\theta}_{4}^{\prime}}{\ov{\theta}_{4}}\right)^{\prime} \;=\; \nu_{\rm t}\;\frac{\ov{\theta}_{4}^{\prime\prime}(0)}{
\ov{\theta}_{4}(0)} \;-\; \frac{\kappa^{-1}}{\nu_{\rm t}}\;\ov{\rm sn}^{2}, \]
and
\[ \nu_{\rm t}\;\frac{\ov{\theta}_{4}^{\prime\prime}(0)}{\ov{\theta}_{4}(0)} \;=\; -\;\nu_{\rm t}^{-1}\left( \frac{\ov{\sf E}}{\ov{\sf K}} \;-\; 1 
\right). \]
By using $p_{\|}R_{\|}\,\partial\vartheta/\partial\zeta_{\rm t}$, given by Eq.~\eqref{eq:ps_zeta_t}, we recover Eq.~\eqref{eq:S_zeta}
\begin{equation}
\pd{S_{\rm t}}{\zeta_{\rm t}} \;\equiv\; -\;J_{\rm t} \;+\; p_{\|}\,R_{\|}\;\pd{\vartheta}{\zeta_{\rm t}}.
\end{equation}
Using Eq.~\eqref{eq:S_Z_t}, we also recover Eq.~\eqref{eq:S_J}
\begin{eqnarray}
\pd{S_{\rm t}}{J_{\rm t}} & = & 2\,\nu_{\rm t}\;\frac{\partial^{\prime}(\sqrt{\kappa}\;\ov{\rm Z})}{\partial\kappa} \label{eq:S_J_proof_t} \\
 & = & \frac{\nu_{\rm t}}{\sqrt{\kappa}} \left[ \ov{\rm Z} \;-\; \left( \frac{-\,\ov{\rm cn}^{2}\,\ov{\rm Z}}{(\kappa - 1)} \;+\; 
\frac{\ov{\rm cn}\,\ov{\rm sn}\,\ov{\rm dn}}{(\kappa - 1)} \right) \right] \nonumber \\
 & = & -\;\frac{\nu_{\rm t}}{\sqrt{\kappa}}\;\frac{\ov{\rm cn}\,\ov{\rm sn}\,\ov{\rm dn}}{(\kappa - 1)} \;+\; \frac{\nu_{\rm t}}{\sqrt{\kappa}}\;\ov{\rm Z}
\left[ 1 \;+\; \frac{\kappa^{-1}\,\ov{\rm cn}^{2}}{(1 - \kappa^{-1})} \right] \nonumber \\
 & = & 2\,\nu_{\rm t}\,\sqrt{\kappa}\,\frac{\ov{\rm dn}}{\ov{\rm cn}}\;\frac{\partial^{\prime}\ov{\rm sn}}{\partial\kappa} \;\equiv\; p_{\|}\,R_{\|}\;
\pd{\vartheta}{J_{\rm t}}
\nonumber
\end{eqnarray}
where we used Eq.~\eqref{eq:sn_kappa}.


\begin{thebibliography}{5}
\bibitem{footnote_1} We note that our definition of the bounce-transit parameter \eqref{eq:kappa} is the square of the parameter used in previous works \cite{Wang_Hahm}. Our definition is consistent with the functional dependence of the Jacobi elliptic functions and the complete elliptic integrals defined within {\sf Mathematica}.
\bibitem{footnote_2} The trapped-particle solutions \eqref{eq:p_b}-\eqref{eq:theta_b} are obtained from Eqs.~(23)-(24) of Ref.~\cite{Brizard_gc_tok} by translation identities ${\rm sn}(\chi_{\rm b} + {\sf K}|\kappa) = {\rm cd}(\chi_{\rm b}|\kappa)$ and ${\rm cn}(\chi_{\rm b} + {\sf K}|\kappa) =
 -\;\sqrt{1 - \kappa}\;{\rm sd}(\chi_{\rm b}|\kappa)$.
\end{thebibliography}
\end{document}